\documentclass{epl}

\title{A new vibrational level of the H$_2^+$ molecular ion}
\author{
J. Carbonell\inst{1},
R. Lazauskas\inst{1},
D. Delande\inst{2},
L. Hilico\inst{2,3}
\and S. K{\i}l{\i}\c{c}\inst{2,3}}
\institute{
\inst{1} Institut des Sciences Nucl\'eaires
         - 53 Av. des martyrs, 38026 Grenoble, France     \\
\inst{2} Laboratoire Kastler Brossel
         - Case 74, 4 place Jussieu, 75252 Paris, France  \\
\inst{3} Universit\'e d'Evry Val d'Essonne
         - Boulevard F. Mitterrand, 91025 Evry cedex, France}

\begin{document}

\maketitle

\begin{abstract}
A new vibrational level of the H$_2^+$ molecular ion
with binding energy of $1.09\times 10^{-9}$a.u.$\approx30$~neV below the first dissociation
limit is predicted, using highly accurate numerical nonrelativistic quantum
calculations, which go beyond the Born-Oppenheimer approximation.
It is the first excited vibrational level $v=1$ of the
2p$\sigma_u$ electronic state, antisymmetric with respect to the
exchange of the two protons, with orbital angular momentum $L=0.$
It manifests itself as a huge p-H scattering length of $a=750\pm 5$ Bohr radii.
\end{abstract}

The H$_2^+$ molecular ion was one of the first non-trivial quantum
mechanical systems
treated in the early days of Quantum Mechanics \cite{Gutzwiller_90}.
Since, its study has been constantly pursued and the bound state
calculations reach nowadays a very high degree of
accuracy \cite{woln, moss,taylor,frolov,korobov,HBGL_EPJD_00}.

The Born-Oppenheimer approximation
provides us with a simple, although approximate, description of the system.
Two electronic energy curves are correlated with the first dissociation limit:
the 1s$\sigma_g$ curve with a rather deep energy well
supporting twenty $L=0$ vibrational states ($L$ is the orbital angluar momentum)
and the 2p$\sigma_u$
which is mainly repulsive.
The later one presents however at large distances a weakly attractive potential whose depth is
about $6\times 10^{-5}$ a.u.
A single $L=0$ level ($v=0$)  with binding energy $1.56625\times 10^{-5}$
a.u. has been found in this well up to now.

We will show in this letter that an
excited  $L=0$ level ($v=1$) of the 2p$\sigma_u$ state of H$_2^+$ actually exists.
Its binding energy turns out to be extremely small -- $1.09 \times
10^{-9}$~a.u., i.e. 30~neV or 0.00024~cm$^{-1}$ -- and the corresponding
wave function has a spatial extension of several hundred Bohr radii, a noticeable
fraction of a micrometer.
This state manifests itself in a huge p-H scattering length  $a=750$~a.u. which dominates
the low energy scattering cross section of proton by atomic hydrogen.
The H$_2^+$ formation rate, as well as the subsequent abundance of H$_2$ molecules,
can be substantially influenced by this resonant p-H cross section. 
The possible existence of
an excited vibrational  $L=0$ level was questioned in~\cite{moss}

Calculations have been performed in the framework of non relativistic
dynamics and in two different ways.
On the one hand, by solving the Faddeev equations
\cite{LC_FBS_02}, what provides the low energy p-H scattering parameters
and -- by mean of the extended effective
range expansion \cite{MERT} -- allows to determine the binding
energies of near-threshold H$_2^+$ states.
On the other hand, by using {\it adhoc} bound state methods~\cite{HBGL_EPJD_00}
which provide the best existing binding energies of the H$_2^+$ system.
Despite their radical differences, in their formal as well
as in their practical implementations,
these two approaches well agree in their bound state predictions.

We summarize in what follows the scattering and bound state results
obtained by these two methods.
As our calculations do not involve Born-Oppenheimer approximation,
we use only exact quantum numbers in  non-relativistic dynamics, where
spin-orbit and spin-spin couplings are neglected (their effects
are discussed below) i.e.:
the total orbital angular momentum $L$, the parity $\Pi$ and the
symmetry property of the spatial wave function with respect to the two-protons exchange;
for the latter symmetry, the symmetric states
have a total proton spin $S_{pp}=0$ (singlet)  and the antisymmetric
ones have $S_{pp}=1$ (triplet).
The connection between exact and approximate quantum numbers
is discussed in~\cite{HBGL_EPJD_00}: for $L=0,$
the 1s$\sigma_g$ levels
are singlet, even parity states, while the 2p$\sigma_u$
levels are triplet, even parity states.

The 3-body (p,p,e$^-$) Faddeev calculations are performed in
configuration space \cite{LC_FBS_02}. Three sets of Jacobi
coordinates, corresponding to the different asymptotic
states, are involved  and defined by
\[\mathbf{x}_{\alpha}=\sqrt{\frac{2m_{\beta}m_{\gamma}}{m_{\beta}+m_{\gamma}}}
(\mathbf{r}_{\beta}-\mathbf{r}_{\gamma}) \qquad
\mathbf{y}_{\alpha}=\sqrt{\frac{2m_{\alpha}(m_{\beta}+m_{\gamma})}
{m_{\alpha}+m_{\beta}+m_{\gamma}}}
(\mathbf{r}_{\alpha}-\frac{m_{\beta}\mathbf{r}_{\beta}+m_{\gamma}\mathbf{r}_{\gamma}}{m_{\beta}+m_{\gamma}})\]
where $(\alpha\beta\gamma)$ denote cyclic permutations of (123), $m_{\alpha}$ the particle
masses and we identify 1$\equiv$p, 2$\equiv$p, 3$\equiv$e$^-$.
The standard Faddeev  equations read
\begin{equation}\label{FE}
(E-H_{0}-V_{\alpha})\Psi_{\alpha}=V_{\alpha}\sum_{\alpha\neq\beta}\Psi_{\beta},
\end{equation}
where $H_{0}$ is the 3-particle free hamiltonian and
$V_{\alpha}$ the 2-body Coulomb potential for the interacting $(\beta\gamma)$ pair.

Equations (\ref{FE}) provide satisfactory solutions for bound states
but are not suitable for scattering Coulomb problems.
The reason is that their right hand side does not decrease fast enough
to ensure the decoupling of the Faddeev amplitudes in the asymptotic region
and to allow  unambiguous implementation of boundary conditions.
In order to circumvent this problem, Merkuriev \cite{Stas_80} proposed to split
the Coulomb
potential $V$ into two parts by means of some arbitrary cut-off function $\chi$
\begin{eqnarray*}
V(\mathbf{x})     &=& V^s(\mathbf{x},\mathbf{y}) +V^l(\mathbf{x},\mathbf{y})  \\
V^s(\mathbf{x},\mathbf{y}) &=& V(\mathbf{x})\chi(\mathbf{x},\mathbf{y}) \\
V^l(\mathbf{x},\mathbf{y}) &=& V(\mathbf{x})[1-\chi(\mathbf{x},\mathbf{y})]
\end{eqnarray*}
and to keep in the right hand side of equation (\ref{FE}) only the short
range $V^s$ contribution.
One is then left with a system of equivalent equations
\begin{equation}\label{MFE}
(E-H_{0}-W-V^s_{\alpha})\bar\Psi_{\alpha}=V^s_{\alpha}\sum_{\alpha\neq\beta}\bar\Psi_{\beta},
\end{equation}
in which $W$ is a 3-body potential containing the long range parts:
\begin{eqnarray*}
W &=& V^l_{\alpha}+V^l_{\beta}+V^l_{\gamma}
\end{eqnarray*}
 The systems of equations (\ref{FE}) and (\ref{MFE})
are strictly equivalent to the Schr\"odinger equation 
and realize two different partitions of the total three-body wave function
\[ \Psi = \Psi_1+\Psi_2+\Psi_3 = \bar\Psi_1+\bar\Psi_2+\bar\Psi_3 \]
This approach was found to be very efficient and accurate in calculating the
low energy e$^+$Ps and e$^+$H cross sections \cite{KCG_PRA}.

Equations (\ref{MFE}) were solved
by expanding amplitudes $\bar\Psi_i$ in the bipolar harmonics basis
\begin{equation}\label{PW}
\bar\Psi_{\alpha}(\mathbf{x}_{\alpha},\mathbf{y}_{\alpha})=\!\!\!
\sum_{\!\!\!i_{\alpha}\equiv\left\{l_{\mathbf{x}_{\alpha}},l_{\mathbf{y}_{\alpha}}\right\}}
\frac{1}{x_{\alpha} y_{\alpha}}
\;{\varphi_{{\alpha},i_{\alpha}}(x_{\alpha},y_{\alpha})}\;
 B_{i_{\alpha}}^{LM}(\hat{\mathbf{x}}_{\alpha},\hat{\mathbf{y}}_{\alpha})
\end{equation}
and  their reduced components $\varphi_{{\alpha},i_{\alpha}}$ in the basis of two-dimensional splines.
Some care has to be taken in extracting the scattering observables,
specially at zero energy, from the asymptotic solution at finite distance.
The long range polarization force -- generated by the equations themselves -- makes the
convergence of the observables as a function of the p-H distance
very slow and requires an appropriate extrapolation procedure. The
first results of these calculations can be found in \cite{LC_FBS_02} and a
detailed explanation of the method  will be given in a forthcoming publication \cite{LC_PRA_02}.

For the 1s$\sigma_g, L=0$ state, we obtain the
scattering length  $a=-29.3$~a.u.
The zero energy Faddeev amplitude has 20 nodes in $y_1$-coordinate,
indicating the existence of 20 vibrational energy levels for H$_{2}^{+}$.

In the  2p$\sigma_u, L=0$ state, our calculations gives $a=750\pm 5$~a.u.
The sign and the nodal structure of the Faddeev amplitudes indicates
that such a big value is due to the existence of a
first excited state with extremely small binding energy.
By calculating the p-H phase shifts at small energies
and using the effective range theory we are able to determine its binding energy.
In presence of long range p-H polarization potential,
the effective range expansion has the form \cite{MERT}
\begin{equation}\label{MERT}
k\cot \delta = -\frac{1}{a} + a_1 k + a_2 k^2 \log k + a_3 k^2 + o(k^2)
\end{equation}
where $k$ is the wave vector and $a_i$ are coefficients depending on the interaction.
It turns out that, in the presence of a weakly bound state with imaginary momentum
$k=ik_0$,
the $a_1$ and $a_2$ terms in (\ref{MERT})  become negligible and  the expansion recovers
the standard form \cite{MERT,MM_87}:
\begin{equation}
k\cot \delta = \alpha + \beta k^2
\end{equation}
with $\alpha=-k_0+\beta k_0^2$.
Coefficients $\alpha$ and $\beta$ are obtained by fitting the phase-shift
and they determine the $k_0$ value.
Using $B=\frac{k_0^2}{m_p}$,
we found \cite{LC_FBS_02} by this procedure a bound state at
$B=1.13\pm0.04\times 10^{-9}$ a.u. below the first p+H dissociation threshold.
To our knowledge, this is the weakest bond ever predicted,
three times smaller than the $^4$He atomic dimer \cite{Dimer}.

\begin{figure}[ht]
\begin{center}
\includegraphics[width=8.cm]{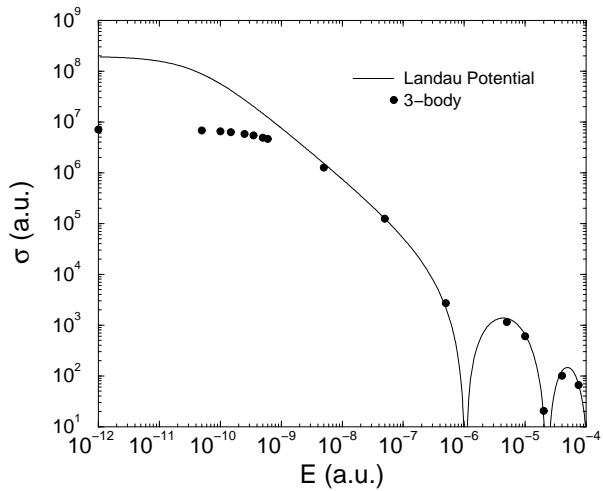}
\caption{Low energy cross section for the scattering of a
proton by an hydrogen atom in its ground state, for the pp triplet state,
compared to the results using the Landau potential for the interaction.}\label{sigma_pH_A}
\end{center}
\end{figure}

The  S-wave p-H  cross section for the triplet ($2p\sigma_u$)  state
is displayed in fig.~\ref{sigma_pH_A} (dotted values).
The  singlet ($1s\sigma_g$) contribution is negligible in the zero energy region.
It is interesting to compare the 3-body calculations with those (solid line)
provided by the simple Landau's two-body potential $V(r)=2r\exp{(-r-1)}-9/4r^4$ \cite{LL_3}.
This model -- based on Pauli repulsion
between protons  overbalanced at $r\sim$10~a.u.
by attractive polarization forces --
gives quite a good result for the ground state $(B=1.4531 \times 10^{-5}$~a.u.
instead of the exact $1.56625\times10^{-5}$ value \cite{HBGL_EPJD_00})
but differs strongly in the predictions of the first excited one $(B=4.0566 \times 10^{-11}$~a.u.).
In the zero energy scattering, both calculations differ also by more than one order
of magnitude while at energies $E\sim$10$^{-8}$
they are already in quite a good agreement.

\begin{figure}[ht]
\begin{center}
\includegraphics[width=6cm,angle=-90]{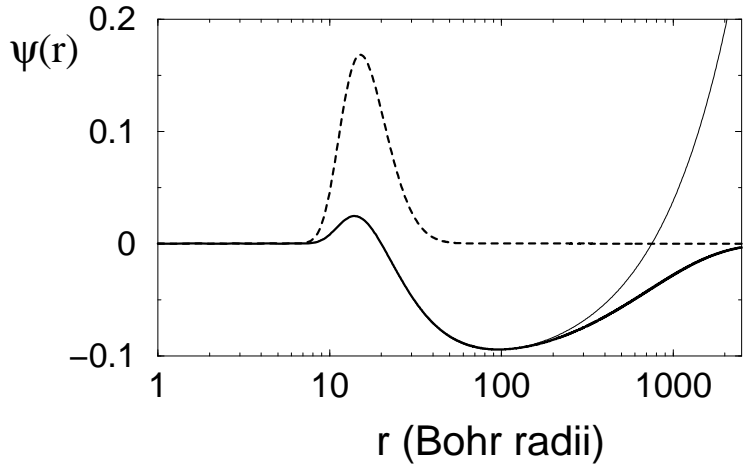}
\caption{Wavefunctions (not normalized) of the ground (dashed line)
and excited (solid thick line) levels of the 2p$\sigma_u$ state
of the H$_2^+$ molecular ion.
The last one is compared with the corresponding p-H zero energy scattering wavefunction
(thin line).
The existence of an excited level with very small binding energy
is predicted by our calculations.
Its wavefunction extends very
far in the internuclear distance $r,$ with a maximum probability
density around 100 Bohr radii. It is responsible for
a huge scattering length of 750 a.u.
Note the use of a logarithmic scale on $r.$}\label{wfc}
\end{center}
\end{figure}

The positive sign of the p-H
scattering length and the existence of one node in the Faddeev amplitude (see  fig.~\ref{wfc})
unambiguously shows the existence of an excited bound state. After this calculation was done, we decided to
use a direct method in order to obtain a more accurate value of the binding
energy as well as a direct computation of its wavefunction.
Because the three particles are bound, the wavefunction must decrease exponentially
if any of the three inter-particles distances goes to infinity. The idea
is thus to expand the full 3-body wavefunction on a convenient discrete
basis set and to diagonalize the 3-body Hamiltonian in this basis
set. For highly accurate calculations of very weakly bound states, the basis
set must be chosen carefully. The first step is to isolate
the angular dependance of the 3body wavefunction, which is straightforward
for $L=0$ states. One is left with a 3-dimensional Schr\"odinger equation
depending on the inter-particle distances only.
We use the perimetric coordinates
\begin{equation}
\begin{array}{c}
x=r_{1}+r_{2}-r_3,\\
y=r_{1}-r_{2}+r_3,\\
z=-r_{1}+r_{2}+r_3,\\
\end{array}
\end{equation}
and express the Schr\"odinger equation as a generalized eigenvalue problem
for the energy $E$:
\begin{equation}
A|\Psi\rangle=E\ B\ |\Psi\rangle,\label{schro-gen}
\end{equation}
where $A$ and $B$ operators are polynomials in the $x,y,z,
\partial_x,\partial_y,\partial_z$ operators.
The basis functions used in the calculation are direct products of
Laguerre polynomials and exponentials along each perimetric
coordinate, whose properties
are discussed in details in \cite{HBGL_EPJD_00}.
The matrices representing $A$ and $B$ in such a basis set are
real symmetric sparse banded matrices, where all
matrix elements are known in analytic form and involve
only simple algebraic expressions. The generalized eigenvalue problem
is then solved using the Lanczos algorithm in order to produce
few eigenvalues, among several thousand, in the interesting energy range.

Whereas an accurate computation of the ground $v=0,L=0$ level of
the 2p$\sigma_u$ state
requires only a moderately large basis set (about 20,000), the computation
of the first excited state is much more difficult and requires at least a basis
size of 150,000 and a careful choice of the variational parameters of the
basis. We used basis sizes up to 450,000 to confirm that
the results discussed here are fully converged. 
For the first excited vibrational level we find a total energy
$E=-0.499727840801511$~a.u. and a dissociation energy
$E=-0.499727839716466$~a.u. This gives a binding energy
$B=1.085045\times 10^{-9}$~a.u. with an  uncertainty of the order
of $10^{-15}$~a.u., due  to numerical precision rather than to the
basis size.  This value is consistent with the one obtained by the
scattering method, but more accurate.

We have also calculated the wavefunctions. These are
full three-body wavefunctions and are thus not easily plotted.
However, they take significant values only for rather
large internuclear distances: the electronic wavefunction is thus essentially
the ground state of the hydrogen atom attached to one of the protons,
independently of the internuclear distance. We checked that
this simple property is almost exactly obeyed by the three-body
wavefunction. Once this trivial part of the wavefunction
is factored out, one is left with a wavefunction depending only on the internuclear
distance. It is plotted in fig.~\ref{wfc} for the lowest two levels.
Because of the very large
size of the excited state, we chose to plot the wavefunction using a
logarithmic scale for the internuclear distance $r.$
The ground level is a nodeless wavefunction centered around $r=15$~a.u.,
while the excited level extends much further;
it has a maximum at $r\approx100$~a.u. and still significant values
at $r\sim1000$~a.u. There is an inflexion point
at $r\approx215$~a.u., located at the outer turning point
of the Born-Oppenheimer (or Landau) potential, where the 2p$\sigma_u$
potential equals the binding energy. In the same figure, we also plot
the zero-energy wavefunction got from the Faddeev approach. At small $r,$ it is
remarkably similar to the wavefunction of the excited level, which is not surprising
considering the very small energy difference.
At large distance, the zero-energy wavefunction diverges linearly
and has a zero at $r\approx a$, the scattering length value 750 a.u.

Results presented above have been obtained with
the mass ratio $m_p/m_e=1836.152701$, as recommended in \cite{codata86}.
Using the more recent 1998 CODATA value \cite{codata98} would not change anything significantly.
Our method makes  possible to compute the energy levels
for any mass ratio of the particles. When the mass ratio
is decreased, the binding energy also decreases until
a critical value beyond which the excited level disappears
and one is left with a single bound state. We estimate this
critial mass ratio to be around 1781. This is rather close to the
actual value which explains why the state is so weakly bound.
The closeness of the critical mass ratio also
explains why the Landau potential discussed above
-- although it is fairly accurate -- gives a wrong
binding energy.

Calculations presented here were performed using a fully non relativistic dynamics
with just Coulomb pair-wise interactions taken into account.
In view of the extreme sensitivity of these results,
it is necessary to quantify the possible relativistic effects.

\begin{figure}[ht]
\begin{center}
\includegraphics[width=8cm]{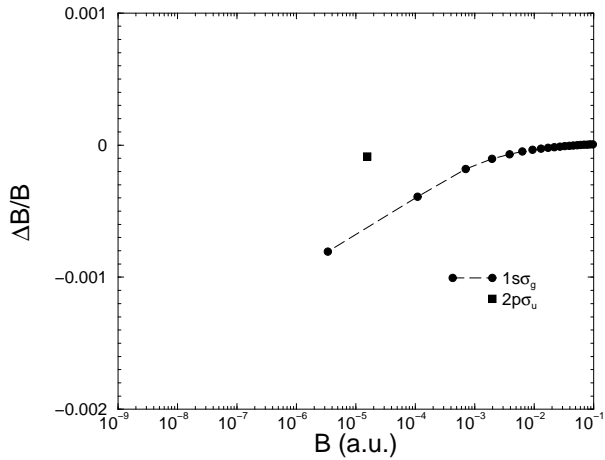}
\caption{Relativistic corrections to the binding energies
for the $L=0$  1s$\sigma_g$ (circles) and the ground 2p$\sigma_u$ (square) state
of the H$_2^+$ molecular ion.}\label{RC_B_H2P}
\end{center}
\end{figure}

As the dynamics is governed by a shallow potential well 
at distances around 10~a.u., where
the nuclear motion is very slow, relativistic corrections must be considered
only for the electronic motion. As the internuclear distance is much larger
than the typical electron-nucleus distance, 
 relativistic corrections
will be essentially identical for the weakly bound state and the
dissociation limit, leaving the binding energy only very weakly modified.
The first order relativistic and radiative
corrections for the H$_2^+$ states have been
obtained in \cite{HK_JCHSF_90} and are discussed with some detail in \cite{moss}.
The results summarizing  the
relative corrections to the binding energy ($\Delta B/B$) for the $L=0$
1s$\sigma_g$ (filled circles)
and 2p$\sigma_u$  (filled square) states are given in fig.~\ref{RC_B_H2P}.
For  1s$\sigma_g$ states  they are smaller than $10^{-3}$  and vary smoothly
over five decades of binding energy. 
For the $v=0$ 2p$\sigma_u$ level they are one order of magnitude smaller.
In absolute value they are $\Delta B\sim10^{-9}$ a.u. 
and as they
 scale as $1/r^4$ \cite{HK_JCHSF_90},
they should be at least one or two orders of magnitude smaller for $v=1$.
Thus, despite the smallness of its binding energy,
relativistic effects preserve the bound character of the new state we have presented.

Another source of relativistic corrections, not included in Refs.
\cite{HK_JCHSF_90,moss}, is the modification of the
$1/r^4$ polarization potential at very large distances due to retardation effects (Casimir-Polder effect).
Contrary to the dipole-dipole case (Van der Waals forces), where
the $1/r^6$ long-range behaviour is changed into $1/r^7$, in the charge-dipole case we are
considering the usual $1/r^4$ term is modified by adding a $1/r^5$ contribution \cite{BT_APHYS_76} in the
form:
\begin{equation}
V(r) = - \frac{\alpha_d}{2r^4} \left(1 - \frac{11\alpha}{2\pi}\frac{m_e}{m_p} \frac{1}{r}\right)
\end{equation}
where $\alpha$ is the fine structure constant. At $r\sim 100$ a.u., the
correction turns out to be negligible.

One should be careful with
the spin-orbit and spin-spin interactions. As all states considered here
have zero total angular momentum $L$, the spin-orbit coupling vanishes
and all states correspond to $J=1/2$ (where ${\bf J}$ is the sum of the
total angular momentum ${\bf L}$ and the electronic spin ${\bf S}.$
The spin-spin interaction is more tricky.
Indeed, we are interested in triplet states with total two-proton spin $S_{pp}=1$,
so that the total angular momentum can be either $F=1/2$ or $F=3/2.$ 
This hyperfine structure
should be very close to the $F=0~/~F=1$ hyperfine structure of the
hydrogen atom. As the latter is much larger than the binding energy we have
calculated for the non-relativistic problem, it is likely that the
$v=1,L=0,J=1/2,F=3/2$ level lies above the
dissociation limit of the $F=1/2$ series. The dissociation
rate induced by the hyperfine coupling is however most probably very low.

A direct measurement of the p-H cross section at very low energy
seems unlikely at present.
One can however access the low energy p-H continuum
in the final state of the H$_2^+$
photodissociation cross section.
The excited vibrational 2p$\sigma_u (v=1, L=0)$ level predicted here is
radiatively coupled to the 1s$\sigma_g (v=19,L=1)$ level. The electric
dipole transition between those two levels should be observable
in the 6~GHz range using an
experiment similar to the one used to detect the
$(v=0,L=0)\rightarrow(v=19,L=1)$ transition \cite{carrington,critchley}.
An experimental confirmation of our results would be very interesting.

\acknowledgments
The authors are sincerely grateful to C. Gignoux, N. Billy and B. Gr\'emaud
for useful discussions and helpful advices.
The numerical calculations were performed  at CGCV (CEA Grenoble) and  IDRIS (CNRS).
We thank the staff members  of these organizations for their constant support.
Laboratoire Kastler-Brossel de l'Universit\'e Pierre et Marie Curie et
de l'Ecole Normale Sup\'erieure is UMR 8552 du CNRS.

\end{document}